\documentclass{article}
\usepackage{amsmath}
\usepackage{cite}
\usepackage{graphicx}
\usepackage{dcolumn}

\begin{document}

\date{}
\title{On some potentials based on exponential functions}
\author{Francisco M. Fern\'{a}ndez\thanks{%
fernande@quimica.unlp.edu.ar} \\
INIFTA, DQT, Sucursal 4, C. C. 16, \\
1900 La Plata, Argentina}
\maketitle

\begin{abstract}
By means of a simple and systematic normalization method we show that some
apparently different potentials based on exponential functions are
equivalent. Present normalization method only requires that the potential
exhibits a minimum and is finite when the radial variable tends to infinity.
\end{abstract}

\section{Introduction}

\label{sec:intro}

In the last years molecular potentials based on exponential functions have
proliferated to the extent of producing a zoo of analytical expressions.
Some time ago Boonserm and Visser\cite{BV11} showed the equivalence of many
potentials in their study of scattering processes. We have recently done
something similar by means of a systematic normalization method with the
focus on bound states\cite{F26}.

The purpose of this paper is the analysis of two M\"{o}bius-square-based
potentials proposed recently\cite{OOEIH20,YSSA24,NJOETAL22,ONYE22}. In
section~\ref{sec:method} we develop a variant of the normalization method introduced in Ref.\cite{F26} that facilitates their
comparison. In section~\ref{sec:Mobius} we apply the normalization method to
the M\"{o}bius square potential, in section~\ref{sec:MobHult} to a
five-parameter model and in section~\ref{sec:general} to a a six-parameter
model. Finally, in section~\ref{sec:conclusions} we summarize out main
results and present out conclusions.

\section{Normalization method}

\label{sec:method}

Throughout this paper we resort to the following\newline
\textbf{Definition}: Two radial potentials $U(r)$ and $V(r)$ are said to be
equivalent if $\Delta =U(r)-V(r)$ is independent of $r$.\newline
Note that the Schr\"{o}dinger equations with $U(r)$ and $V(r)$ exhibit the
same eigenfunctions and the eigenvalues are related by $%
E_{nl}^{U}=E_{nl}^{V}+\Delta $, $n,l=0,1,\ldots $. Besides, if $V_{1}(r)$ is
equivalent to $V_{2}(r)$ and the latter is equivalent to $V_{3}(r)$ then $%
V_{1}(r)$ is equivalent to $V_{3}(r)$. For this reason we can speak of
classes of equivalent potentials. In order to prove the equivalence of some
given potential-energy functions we resort to a simple normalization method
derived from our earlier procedure\cite{F26}.

Consider a potential $V(r)$, $r\geq 0$, that depends on more than two
adjustable parameters. We require that this potential exhibits a minimum at $%
r=r_{e}$ and a finite value when $r\rightarrow \infty $ (like most commonly
accepted molecular potentials). In order to achieve this goal we force the
given analytical function to satisfy the equations
\begin{equation}
V^{\prime }\left( r_{e}\right) =0,\;\lim\limits_{r\rightarrow \infty
}V(r)-V\left( r_{e}\right) =D_{e}>0,  \label{eq:re,De}
\end{equation}
from which we obtain two model parameters in terms of $r_{e}$, $D_{e}$ and
the remaining potential parameters. We then rewrite the potential in the
normal form\cite{F26}
\begin{equation}
V(r)-V\left( r_{e}\right) =D_{e}\left[ 1-f(r)\right] ^{2}.  \label{eq:V_f}
\end{equation}
Since $\lim\limits_{r\rightarrow \infty }\left[ 1-f(r)\right] ^{2}=1$ we
conclude that the two roots $f_{0}(r)$ and $f_{2}(r)$ of the quadratic
equation (\ref{eq:V_f}) behave asymptotically as $\lim\limits_{r\rightarrow
\infty }f_{0}(r)=0$ and $\lim\limits_{r\rightarrow \infty }f_{2}(r)=2$. We
always choose $f_{0}(r)$ in order to provide a consistent representation of $%
V(r)$ in normal form. According to the definition given above two potentials
with the same normal form are obviously equivalent.

As an example, consider
\begin{equation}
V(r)=V_{0}+\frac{V_{1}}{e^{\alpha r}+a}+\frac{V_{2}}{\left( e^{\alpha
r}+a\right) ^{2}},\;\alpha >0.  \label{eq:V_example}
\end{equation}
Clearly, $V_{0}$ is an irrelevant energy shift that does not appear in
equations (\ref{eq:re,De}). If we solve equations (\ref{eq:re,De}) for $%
V_{1} $ and $V_{2}$ we obtain
\begin{equation}
V_{1}=-2D_{e}\left( e^{\alpha r_{e}}+a\right) ,\;V_{2}=D_{e}\left( e^{\alpha
r_{e}}+a\right) ^{2},  \label{eq:V1,V2_example}
\end{equation}
and the normal form
\begin{equation}
V(r)-V\left( r_{e}\right) =D_{e}\left( 1-\frac{e^{\alpha r_{e}}+a}{e^{\alpha
r}+a}\right) ^{2},  \label{eq:V-Ve_example}
\end{equation}
of the potential (\ref{eq:V_example}).

\section{M\"{o}bius square potential}

\label{sec:Mobius}

Boonserm and Visser\cite{BV11} showed that several potentials are equivalent
to the M\"{o}bius square potential
\begin{equation}
V(r)=V_{0}\left( \frac{A+Be^{-\alpha r}}{C+De^{-\alpha r}}\right) ^{2},
\label{eq:V_Mob}
\end{equation}
that was also used by Ikot et al.\cite{IYZH14} in their calculations. Note
that this potential can be rewritten as
\begin{equation}
V(r)=U_{0}\left( \frac{1+be^{-\alpha r}}{1+de^{-\alpha r}}\right)
^{2},\;U_{0}=V_{0}\frac{A^{2}}{C^{2}},\;b=\frac{B}{A},\;d=\frac{D}{C},
\label{eq:V_Mob_2}
\end{equation}
which clearly shows that two parameters are redundant and can be set equal
to unity without loss of generality (we can alternatively set to unity $B$
and $D$ or $B$ and $C$ or $A$ and $D$). Here we choose $A=C=1$ from now on.
This redundancy was useful for the purposes of Boonserm and Visser\cite{BV11}
but Ikot et al.\cite{IYZH14} gave numerical values to all the parameters as
if they were independent.

Straightforward application of the normalization method yields
\begin{equation}
U_{0}=D_{e},\;b=-e^{\alpha r_{e}},  \label{eq:U0,b_Mob}
\end{equation}
and
\begin{equation}
V(r)=D_{e}\left( 1-\frac{e^{\alpha r_{e}}+d}{e^{\alpha r}+d}\right) ^{2},
\label{eq:V_Mob_3}
\end{equation}
because $V\left( r_{e}\right) =0$. We appreciate that the M\"{o}bius square
potential is equivalent to our example in section~\ref{sec:method}. The
M\"{o}bius square potential exhibits only four independent parameters and
not six as Ikot et al.\cite{IYZH14} appear to believe.

\section{M\"{o}bius square plus Hult\'{e}n potential}

\label{sec:MobHult}

Onyenegecha et al\cite{OOEIH20} and Yanar et al\cite{YSSA24} discussed

\begin{equation}
V(r)=-V_{0}\left( \frac{A+Be^{-2\alpha r}}{1-e^{-2\alpha r}}\right) ^{2}+%
\frac{V_{1}e^{-2\alpha r}}{1-e^{-2\alpha r}},  \label{eq:V_Mob+Hult}
\end{equation}
as if it were a five-parameter potential. However, it can be rewritten as

\begin{equation}
V(r)=-U_{0}\left( \frac{1+be^{-\beta r}}{1-e^{-\beta r}}\right) ^{2}+\frac{%
V_{1}e^{-\beta r}}{1-e^{-\beta r}},\;U_{0}=V_{0}A^{2},\;b=\frac{B}{A}%
,\;\beta =2\alpha ,  \label{eq:V(r)_2}
\end{equation}
which shows that it exhibits at most four independent parameters and that we
can choose $A=1$ without loss of generality.

Upon applying the normalization method we obtain
\begin{equation}
U_{0}=-\frac{D_{e}\left( e^{\beta r_{e}}-1\right) ^{2}}{\left( b+1\right)
^{2}},\;V_{1}=-\frac{2D_{e}\left( e^{\beta r_{e}}-1\right) \left( e^{\beta
r_{e}}+b\right) }{b+1},  \label{eq:V0,V1_Mob+Hult}
\end{equation}
and
\begin{equation}
V(r)-V\left( r_{e}\right) =D_{e}\left( 1-\frac{e^{\beta r_{e}}-1}{e^{\beta
r}-1}\right) ^{2}.  \label{eq:V-Ve_Mob+Hult}
\end{equation}
It is worth noting that $V(r)$ and $V\left( r_{e}\right) $ depend on $b$ but
their difference does not. This fact clearly shows that the effect of $b$ is
merely to produce an arbitrary energy shift and, consequently, it is an
irrelevant parameter (in fact, if we choose $b=0$ we obtain the same
result). Onyenegecha et al\cite{OOEIH20} and Yanar et al.\cite{YSSA24} did
not realize that the potential (\ref{eq:V_Mob+Hult}) does not exhibit five
parameters but only three. Besides, the potential (\ref{eq:V_Mob+Hult}) is a
particular case of (\ref{eq:V_example}) and (\ref{eq:V_Mob}) with $a=-1$ and
$d=-1$, respectively.

We can proceed in a different way and solve equations (\ref{eq:re,De}) for $%
U_{0}$ and $b$, thus obtaining
\begin{equation}
U_{0}=-\frac{\left( 2D_{e}e^{\beta r_{e}}-2D_{e}+V_{1}\right) ^{2}}{%
4D_{e}\left( e^{\beta r_{e}}-1\right) ^{2}},\;b=-\frac{2D_{e}e^{2\beta
r_{e}}-2D_{e}e^{\beta r_{e}}+V_{1}}{2D_{e}e^{\beta r_{e}}-2D_{e}+V_{1}},
\label{eq:U0,b_Mob+Hult}
\end{equation}
and the same normalized potential (\ref{eq:V-Ve_Mob+Hult}). This fact
suggests that we can choose $V_{1}=0$ without loss of generality. In fact,
when $V_{1}=0$ we have
\begin{equation}
U_{0}=-D_{e},\;b=-e^{\beta r_{e}},  \label{eq:U0,b_Mob+Hult_V1=0}
\end{equation}
ane the same potential (\ref{eq:V-Ve_Mob+Hult}). From this result we
conclude that the Hulth\'{e}n potential is irrelevant (contributes only to
an additive energy shift) and that all the physical features of the model
come from the M\"{o}bius one.

\begin{figure}[tbp]
\begin{center}
\includegraphics[width=9cm]{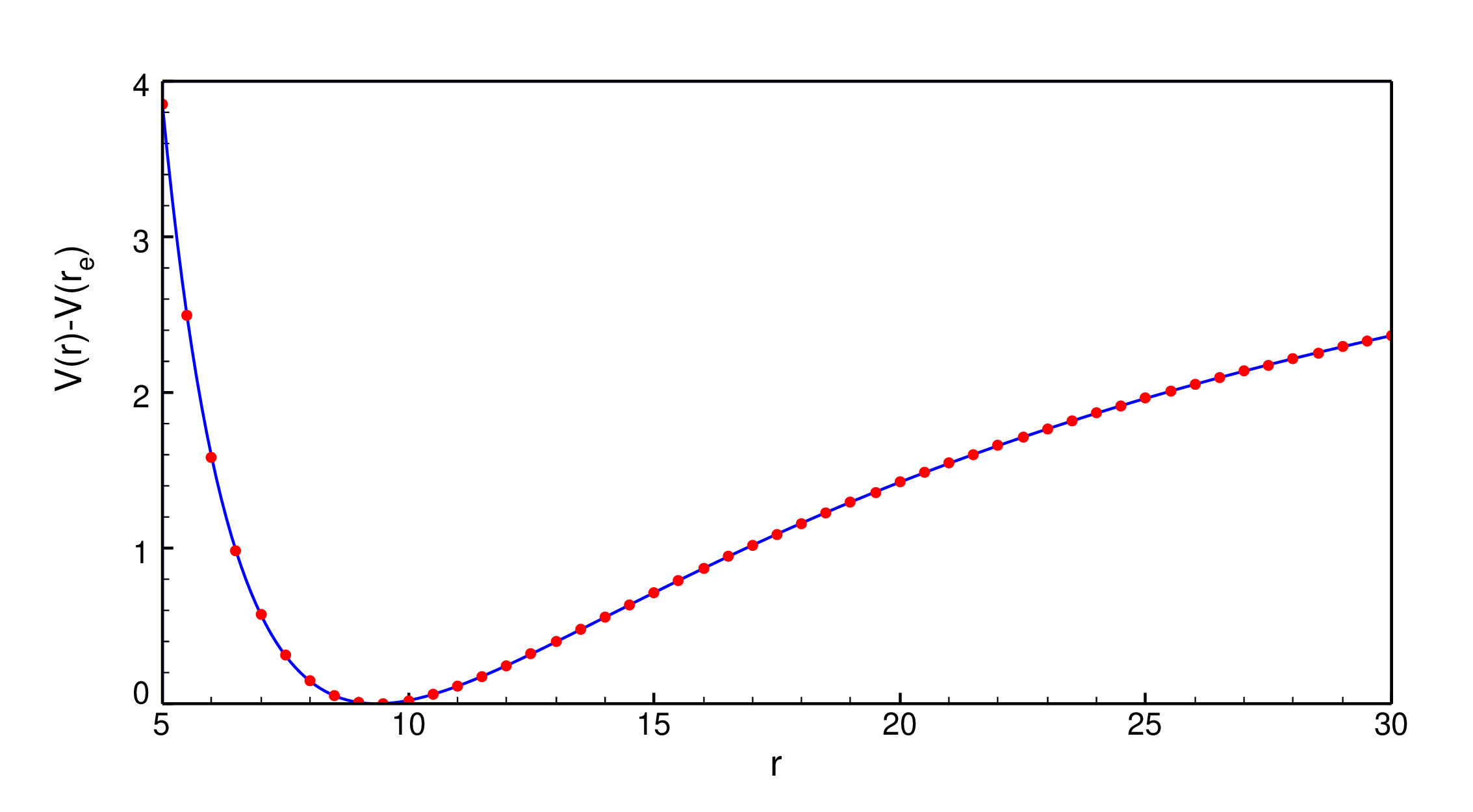}
\end{center}
\caption{$V(r)-V(r_e)$ from equation (\ref{eq:V_Mob+Hult}) with $A=0.9$, $%
B=-1.08$, $V_0=-5.5$, $V_1=0.05$, $\alpha=0.01$ (solid blue line) and the
right-hand side of equation (\ref{eq:V-Ve_Mob+Hult}) with $D_e=4.208507295$,
$r_e=9.356070620 $, $\beta=0.02$ (red circles)}
\label{Fig:V001}
\end{figure}

Yanar et al.\cite{YSSA24} showed some vibration-rotation energies for the
potential (\ref{eq:V_Mob+Hult}) with parameters $A=0.9$, $B=-1.08$, $%
V_{0}=-5.5$, $V_{1}=0.05$ and $\alpha=0.01$. With these parameters we obtain
$D_{e}=4.208507295$ and $r_{e}=9.356070620$. Figure~\ref{Fig:V001} compares $%
V(r)-V\left( r_{e}\right) $ calculated with (\ref{eq:V_Mob+Hult}) and the
parameter values just mentioned (solid, blue line) with the right-hand side
of equation (\ref{eq:V-Ve_Mob+Hult}) and the corresponding values of $D_{e}$
and $r_{e}$ (red circles). We see that the two potentials are identical as
predicted by our analytical equations.

\section{Apparently more general six-parameter representation}

\label{sec:general}

Yanar et al.\cite{YSSA24} mentioned the apparently more general
six-parameter potential-energy function
\begin{equation}
V(r)=-U_{0}\left( \frac{A+Be^{-2\alpha r}}{1-e^{-2\alpha r}}\right) ^{2}-%
\frac{U_{1}e^{-2\alpha r}}{1-e^{-2\alpha r}}+\frac{U_{2}e^{-2\alpha r}}{%
\left( 1-e^{-2\alpha r}\right) ^{2}},  \label{eq:V_Mob+Hult_gen}
\end{equation}
that had been used earlier by Njoku et al.\cite{NJOETAL22} and Onyenegecha
et al.\cite{ONYE22}. In this case we can also choose $A=1$ as argued above.
If we apply the normalization method we obtain
\begin{eqnarray}
U_{0} &=&-\frac{D_{e}\left( e^{\beta r_{e}}-1\right) \left( e^{\beta
r_{e}}\left( U_{2}-1\right) +U_{2}+1\right) }{\left( B+1\right) \left(
B\left( U_{2}-1\right) -U_{2}-1\right) },\;  \nonumber \\
U_{1} &=&-\frac{2D_{e}\left( e^{\beta r_{e}}-1\right) \left( e^{\beta
r_{e}}+B\right) }{B\left( U_{2}-1\right) -U_{2}-1},
\label{eq:U0,U1_Mob+Hult_gen}
\end{eqnarray}
and exactly the result given in equation (\ref{eq:V-Ve_Mob+Hult}).

If we solve equations (\ref{eq:re,De}) for $U_{0}$ ad $B$ we obtain
\begin{eqnarray}
U_{0} &=&-\frac{\left( 2D_{e}e^{2\alpha r_{e}}-2D_{e}+U_{1}\left(
U_{2}-1\right) \right) ^{2}}{4\left( D_{e}e^{4\alpha r_{e}}-2D_{e}e^{2\alpha
r_{e}}+D_{e}-U_{1}U_{2}\right) },\;  \nonumber \\
B &=&-\frac{\left( 2D_{e}e^{4\alpha r_{e}}-2D_{e}e^{2\alpha
r_{e}}-U_{1}\left( U_{2}+1\right) \right) }{2D_{e}e^{2\alpha
r_{e}}-2D_{e}+U_{1}\left( U_{2}-1\right) },  \label{eq:U0,b_Mob+Hult_gen}
\end{eqnarray}
and equation (\ref{eq:V-Ve_Mob+Hult}). This last result enables us to choose
$U_{1}=U_{2}=0$ without affecting the physical content of the model. If we solve
equations (\ref{eq:re,De}) for $U_{1}$ and $U_{2}$ we obtain (\ref
{eq:V-Ve_Mob+Hult}) which tells us that we can choose $U_{0}=0$ without
affecting the performance the predictive capability. We appreciate that the potentials (%
\ref{eq:V_Mob+Hult}) and (\ref{eq:V_Mob+Hult_gen}) are equivalent and
particular cases of (\ref{eq:V_example}) and (\ref{eq:V_Mob}).

\section{Conclusions}

\label{sec:conclusions}

The application of a simple normalization method shows that the
five-parameter potential (\ref{eq:V_Mob+Hult}) and the six-parameter
potential (\ref{eq:V_Mob+Hult_gen}) are equivalent to the three-parameter
potential shown in the right-hand side of equation (\ref{eq:V-Ve_Mob+Hult}).
The conclusion is that the research papers\cite
{OOEIH20,YSSA24,NJOETAL22,ONYE22} are based on the same potential in
different disguises. Here, we have provided what seems to be the minimal
expression for such potential-energy functions.

The normalized form of the potential (\ref{eq:V-Ve_Mob+Hult}) is certainly
simpler than the original potentials that contain redundant parameters. In
addition to it, $D_{e}$, $r_{e}$ and $\alpha $ have an obvious spectroscopic
meaning. The first two parameters are given in most textbooks on
spectroscopy and $\alpha $ can be obtained from the force constant $%
k=V^{\prime \prime }\left( r_{e}\right) $ that is related to the vibrational
frequency\cite{H50}.

The normalization procedure proposed here can therefore serve as a practical
test for identifying redundant parameters and apparently novel molecular
potentials before undertaking analytical or numerical studies. 
Present normalization method maps every member of an equivalence class onto a unique canonical representative.

\begin{thebibliography}{9}
\bibitem{BV11}  P. Boonserm and M. Visser, JHEP \textbf{03}, 073 (2011).

\bibitem{F26}  F. M. Fern\'{a}ndez, Equivalent potential-energy functions,
arXiv:2606.00378 [math-ph].

\bibitem{OOEIH20}  C. P. Onyenegecha, C. A. Onate, O. K. Echendu, A. A. Ibe,
and H. Hassanabadi, Eur. Phys. J. Plus \textbf{135}, 289 (2020).

\bibitem{YSSA24}  H. Yanar, O. Sahin, M. Salti, and O. Aydogdu, Eur. Phys.
J. Plus \textbf{139}, 896 (2024).

\bibitem{NJOETAL22}  I. J. Njoku, E. Onyeocha, C. P. Onyenegecha, M. Onuoha,
E. K. Egeonu, and P. Nwaokafor, Int. J. Quantum Chem. \textbf{123}, e27050
(2022).

\bibitem{ONYE22}  C. P. Onyenegecha, I. J. Njoku, A. I. Opara, O. K.
Echendu, E. N. Omoko, F. C. Eze, C. J. Okereke, E. Onyeocha, and F. U.
Nwaneho, Heliyon \textbf{8}, e08952 (2022).

\bibitem{IYZH14}  A. N. Ikot, B. H. Yazarloo, S. Zarrinkamar, and H.
Hassanabadi, Eur. Phys. J. Plus \textbf{129}, 79 (2014).

\bibitem{H50}  G. Herzberg, Molecular Spectra and Molecular Structure. I.
Spectra of Diatomic Molecules, Second ed. (Van Nostrand Reinhold, New York,
1950).
\end{thebibliography}
\end{document}